\documentstyle [12pt] {article}
\begin{document}
\title {Dark Matter: The Problem of Motion}
\maketitle
\begin{center}
{\bf { Magd E. Kahil }{\footnote{Faculty of Engineering, Modern Sciences and Arts University , Giza , Egypt \\
{{e.mail: mkahil@msa.eun.eg \\ magdelias.kahil@gmail.com}} }}}\footnote{Egyptian Relativity Group. Cairo, Egypt}
\end{center}
\begin{abstract}

{\bf{ Dark Matter maybe regarded  by studying the motion of objects, following non-geodesic trajectories. Whether due to the existence of extra mass as a projection of higher dimensions onto lower ones or motion dipolar particles and fluids at the halos of spiral galaxies. The effect of dark matter has been extended nearby the core of the galaxy, by means of the excess of mass appeared in the motion of fluids in the accretion disc. Non-geodesic equations and their deviation ones are derived in the presence of different classes bi-metric theories of gravity. The stability of these trajectories using geodesic deviation technique has been investigated.}}
\end{abstract}
\section{Introduction}
{\bf{Flat rotational curves for spiral galaxies cannot be explained by Newtonian or Einstein's gravity since neither the former or the latter theories of gravity are satisfied and therefore such deviations from these 2 established theories constitutes the existence of Dark Matter. (DM).   In our galaxy meticulous observations have confirmed that rotational velocities range between 200 -300 km/s provided that these clouds are considered to be moving in circular orbits.\\}}
  Yet, there  are alternative remedies related  to modify Newtonian or Einsteinian gravity by changing their corresponding  gravitational potential $\phi$ to become  $\phi = -GM[1+\alpha \exp{(-r/r_{0})}]/(1+\alpha)$ , such that $\alpha = -0.9$ and $ r_{0} \approx 30 kpc $ , in order to explain the behavior of  the flat rotational curves for spiral galaxies.[1]Consequently, such a unique explanation of discrepancy between theory and observation is still under debate. This may lead us to revisit the notation of DM as to be expressed in terms of a mass excess quantity.

  This type of explanation is represented as the behavior of objects following non-geodesic equations or being shown as the projection of the fifth term on the its other 4 dim components of geodesic equation satisfying space-time-matter (STM) [2] . Also, DM can be explained as the behavior of dipolar particles in the presence of polarization field [3].  However, this type of description has been amended to be regarded as dipolar fluids due to taking the effect of dark energy (DE) in the halos [4].

   Thus, one must take into account that DE occupies $74 \%$ of invisible matter of universe; while DM occupies about $ 23 \%$ of that one [5]. This may gives an indication that DM particles are detected in several regions in the universe beside the halos.

   Accordingly, some incidents like the excess of gamma ray radiation nearby the core of the galaxy can be revealed due to DM annihilation [6]. In this case, the effect of dark matter is observed as an excess of mass in the hydrodynamical equations for the accretion disk [7]. Moreover, it has been considered that cause of slight deviation of perihelion motion is counted due the influence  of dark matter particles [8].

   Now, it is essential to implement the significance of studying the problem of motions for the suspected particles or fluids in order to give a possible scenario for the behavior of DM at different scales in the universe. Accordingly, one must seek an appropriate theory of gravity able to detect its existence at different scales. One of the candidates is studying a class of bi-metric theory of gravity which is able to express strong gravitational fields like SgrA* , neutron stars and binary pulsars. Also,to able to express weak gravitational fields playing the same role of general relativity [9].

From this perspective, it is vital in our study to derive the candidate equations of motion showing the mass excess term is due to the existence of dark matter. Such a vital question should be addressed:  What is dark matter?\\
  In our present work, {\bf{ it can be possible to illustrate its causality by three different rival explanations }}:\\
(I) The existence of a scalar field associated with the Galaxy's gravitational field? [1]\\
(II) The projection of a higher dimension spatial dimension on the 4-dim manifold? [2]\\
(III)  Motion of dipolar particles /fluids as claimed in spiral galaxies? [4]  \\

   Consequently, we are going to deal with expressing, the behavior of  dark matter in terms of  non-geodesic equations, these equations are derived using the Lagrangian formalism using the Bazanski-like Lagrangian [10]. This type of equation may give rise to geometrize all trajectories associated with the appearance of dark matter. In other words, the appropriate path equations as described in the Riemannian geometry to represent dipolar particles or fluids of the halos; and the corresponding path equations that represents the hydrostatic stream of fluids in of accretion disk.due to solving the non-geodesic  deviation equation, we can give rise to examine stability conditions, which means an indication of remaining DM effect on each observed regions.

On the other hand , another approach to reveal the above discrepancies between theory and observation at galactic level is due to modified newtonian gravity (MOND) [11] or its bi-metric version BIMOND [12]. These types of theories are rejecting the existence of dark matter and dark energy [DE] and refer such an anomaoly is due to a deficiency in obtaining an appropriate theory of gravity able to cure the Newtonian explanation . Even though , Blanchet has regarded the MOND as gravitational polarization effect [13].

 From this perspective, we are going to derive to apply the appropriate Bazanski-like Lagrangians [14] to examine the equivalence of non-geodesic trajectories with each of the following equations, dipolar moment and dipolar fluids and hydrostatic stream of motion as described in General Relativity  in section 2 .  We extend the previous equations to be expressed in different versions of bi-metric theories of gravity as shown in sec 3.
Finally, it turns out that, the problem of detecting the existence of DM is connected with studying the behavior of the stream of fluids in different gravitational fields. \\
{This may raise the necessity to examine the stability of these systems for being affected by dark matter. This can be seen,  by solving the different corresponding deviation equation for examining  the stability condition, using an independent method of coordinate transformation [15-16] which is be described in sec 4. }
{\bf{
\section{Dark Matter: Equations of Motion from Different Perspectives  }
\subsection{ Dark Matter: Non-Geodesic Equations}
{\bf{Dark matter can be detected its presence due to the excess of mass as appeared in non-geodesic trajectories}}.  These equations are obtained by applying the Euler-Lagrange equation on the following Lagrangian [1] :
\begin{equation}
L {\stackrel{def.}{=}} m(s) g_{\mu \nu}U^{\mu} \frac{D \Psi^\nu}{Ds} +  m(s),_{\rho}\Psi^{\rho}
\end{equation}
 where $U^{\mu}$ is a unit tangent vector, $\Psi^{\nu}$ its corresponding deviation vector,   $m(s)$  is its mass ,to be considered as function of the parameter, and $\mu=1,2,3,4$. provided that:
\begin{equation}
\frac{d \partial L}{ds \partial{{\dot\Psi^{\alpha}}}} - \frac{\partial L}{ \partial \Psi^{\alpha}} =0
\end{equation}
one gets,
\begin{equation}
\frac{dU^{\alpha}}{ds}+ \Gamma^{\alpha}_{\beta \delta} U^{\beta}U^{\delta} = \frac{m{(s)}_{,\beta}}{m{(s)}} (g^{\alpha \beta}- U^{\alpha}U^{\beta})
\end{equation}

such that
$$
m(s)= - \nabla[g(\psi)\psi],
$$
 where $g(\psi) \psi$, is a scalar function , in which the right hand side of equation (3) behaves as a parallel  force to represent the presence of dark matter.

Also, its corresponding non-geodesic deviation equation is obtained by using the commutation relation on equation (3) i.e.
  $$
  A^{\mu}_{; \nu \rho} - A^{\mu}_{; \rho \nu} = R^{\mu}_{\beta \nu \rho} A^{\beta},
  $$
  where $A^{\mu}$ is an arbitrary vector,$R^{\mu}_{\beta \nu \rho}$ is the curvature tensor .\\
 Multiplying both sides by arbitrary vectors, $U^{\rho} \Psi^{\nu}$ as well as taking into consideration the following condition [15]
 $$
 U^{\alpha}_{; \rho} \Psi^{\rho} =  \Psi^{\alpha}_{; \rho } U^{\rho}.
 $$
 Thus, we obtain the corresponding deviation equations
\begin{equation}
\frac{D^{2}\Psi^{\mu}}{Ds^2}= R^{\mu}_{\nu \rho \sigma}U^{\nu}U^{\rho} \Psi^{\sigma} + (\frac{m{(s)}_{,\beta}}{m{(s)}} (g^{\alpha \beta}- U^{\alpha}U^{\beta}))_{;\rho} \Psi^{\rho}.
\end{equation}

 Yet, for examining  the flat rotation curves, it has been found [2] by taking $ \sigma$ as a parameter describing the trajectories of particles on this region , such that $ s \sim \sigma$ , to  obtain the following relation  \begin{equation}\frac{1}{m}\frac{dm}{d \sigma} \equiv {\sqrt{\Lambda/2}} \end{equation}
in which  to be expressed as,
\begin{equation} \frac{1}{m}\frac{dm}{d \sigma} \approx 2 a_{0}/c^2 \end{equation}

 where $a_{0}$ is a constant of acceleration, {$a_{0} \sim 2 \times 10^{-10} m/sec^2$}, as known  of the MOND and c is the speed of light.

Accordingly, we can find that the non-geodesic  equation can be related to MOND [11] in the following way:
\begin{equation}
\frac{d\hat{U}^{\alpha}}{d\sigma}+ \Gamma^{\alpha}_{\beta \delta} \hat{U}^{\beta}\hat{U}^{\delta} = 2 \frac{ a_{0}}{c^2} \hat{U}_{\beta} (g^{\alpha \beta}- \hat{U}^{\alpha}\hat{U}^{\beta})
\end{equation}
where, $\hat{U}^{\alpha} = \frac{dx^{\alpha}}{d \sigma}$ its associated unit tangent vector. \\
Consequently, its corresponding deviation equation becomes
\begin{equation}
\frac{D^{2}\hat{\Psi}^{\mu}}{D{\sigma}^2}= R^{\mu}_{\nu \rho \sigma}\hat{U}^{\nu}\hat{U}^{\rho} \hat{\Psi}^{\sigma} +  2 \frac{ a_{0}}{c^2}( \hat{U}_{\beta} (g^{\alpha \beta}- \hat{U}^{\alpha}\hat{U}^{\beta}))_{; \rho} \hat{\Psi}^{\rho}
\end{equation}
where $\hat{\Psi^{\mu}}$ its corresponding non-geodesic deviation vector. }}

\subsection{ Dark Matter: An Extra-dimensional Effect}
 It is well known that the non-geodesic equations are expressed as , the four components of a geodesic equations for a test particle [1] in  a non-compact space-time $g_{AB,5} \neq 0$ following Wesson's approach of space-time-matter[2].
Thus, the characteristics of dark matter can be appeared within solving the  geodesic equation in 5-dim., provided that
$$
\frac{dS}{ds}= \sqrt{(1 + \epsilon \hat{\Phi}^2 (U^5)^2 )}
$$
such that $\hat{\Phi}$ is a scalar function, and $\epsilon = \pm 1 $.\\
Thus, it can possible to suggest the following Lagrangian:
\begin{equation}
L= g_{AB} U^{A}\frac{D \Psi^{A}}{DS},
\end{equation}
where $A=1,2,3,4,5$. \\
Thus, taking the variation with respect to $\Psi^{C}$ and $U^{C}$ respectively, one can find\\
(i) Equation of Geodesic:
\begin{equation}
\frac{D U^{C}}{DS}=0,
\end{equation}
(2) Equation of Geodesic Deviation:
\begin{equation}
\frac{D^2 \Psi^{C}}{DS^2} = R^{C}_{BDE}U^{B}U^{D}\Psi^{E}.
\end{equation}
With taking into account that the force appeared on its right hand side is expressed within the component of the fifth dimension  of a 5-dim manifold.
Accordingly, equation (3) may be expressed as
$$
\frac{d^2 x^{\mu} }{d S^{2}} + \Gamma^{\mu}_{AB}\frac{d x^{A}}{dS}\frac{d x^{B}}{dS}=0
$$

$$
\frac{d^2 x^{\mu}}{d S^{2}} + \Gamma^{\mu}_{\mu \nu}\frac{d x^{\mu}}{dS}\frac{d x^{\nu}}{ds}= -(\Gamma^{\mu}_{\mu \nu}\frac{d x^{\mu}}{dS}\frac{d x^{\nu}}{dS} + \Gamma^{\mu}_{\mu \nu}\frac{d x^{5}}{dS}\frac{d x^{5}}{dS}).
$$

Meanwhile, by solving equation (10) and considering its fifth component to be substituted in the other four components, this may be regarded as similar  to the behavior of dark matter particles in (3) .\\

 Thus, we find  that the indication of dark matter may be represented in terms of excess of mass in the right hand side of the non-geodesic equation. Such an equation  is obtained as the projection of the fifth component of the geodesic equation onto its counterpart the four dimensional components.
\subsection{ Dark Matter: Equations of Motion Dipolar Moment Particles in The Halo }
  {\bf{ A rival explanation for the cause of the flat rotational curves of spiral galaxies can be expressed due to  the presence of   dipolar dark matter particles [3]. Such particles are not purely dipolar as the involve monopole contribution from the stress-energy momentum tensor obtained from Einstein field equations. }} It has been proposed by Blanchet et al that these particles are examined in terms of studying their corresponding equations of motion, composed of two system of  equations, one may be described $P^{\mu}$ the (passive) linear momentum vector and  $\Omega^{\mu}$  the evolution vector , describing  microscopic (active) momentum- acting as the spin tensor $S^{\mu \nu}$ in the Papapetrou equation of motion for spinning objects [15].  These equations are obtained using Lagrangian formalism is analogous to the its counterpart the motion of spinning with precession {\footnote {see Appendix A}}.

  Thus, we suggest the following Lagrangian:
\begin{equation}
L{\stackrel{def.}{=}} g_{\alpha \beta} P^{\alpha}\frac{D \Psi_{(1)}^{\beta}}{Ds} + \Omega_{\alpha} \frac{D \Psi_{(2)}^{\beta}}{Ds}   + f_{\alpha}\Psi_{(1)}^{\alpha}+ \hat{f}_{\alpha}\Psi_{(2)}^{\alpha},
\end{equation}
in which
$$
P^{\mu} = (2mU^{\mu} + \frac{D \pi^{\mu}}{D s}),
$$
where $\pi^{\mu}$ is  dipolar vector and $\Psi_{(1)}^{\mu}$ is the non-geodesic deviation from the world line  and $\Psi_{(2)}^{\mu}$ is the evolution deviation due to dipole moment; with taking the raising and lowering indices for the evolution vector is by $h^{\mu \nu}$ the projector tensor i.e.
\begin{equation}
h^{\mu \nu} = g^{\mu \nu}- U^{\mu}U^{\nu},
\end{equation}
$$ \bar{\Omega^{\mu}} = h^{\mu \nu} \Omega_{\nu}.
$$

Taking the variation with respect to $\Phi_{1}^{\mu}$ and $\Phi_{2}^{\mu}$ separately we obtain the following set of equation of motion and evolution respectively:
\begin{equation}
\frac{D P^{\mu}}{D s} = f^{\mu},
\end{equation}
and
\begin{equation}
\frac{D \Omega^{\mu}}{D s} = \hat{f}^{\mu},
\end{equation}
such that{\bf{
$$ f^{\mu}= 2m \frac{\bar{\pi}_{\nu}}{\bar \pi} \frac{d V}{d x}(\frac{\bar \pi}{m}), $$
where,
$\bar{\pi} = h^{\mu \nu} {\pi}_{\nu}$, and $V$ is an associated potential function in terms of dipolar vectors.}}

 While the evolution equation becomes
 \begin{equation}
 \frac{D \bar{\Omega}^{\mu}}{D s} = \hat{f}^{\mu},
 \end{equation}

    provided that $\hat{f}  = R^{\mu}_{\nu \rho \sigma} \hat{\pi}^{\sigma} U^{\rho} U^{\nu}.$

 Similarly, using (A.4) and (A.5) as in [2.1], we obtain the corresponding geodesic deviation equations:

\begin{equation}
\frac{D^2 \Psi_{(1)}^{\mu}}{DS^2}= R^{\mu}_{\nu \rho \sigma}P^{\nu} U^{\rho} \Psi_{(1)}^{\sigma}+ f^{\mu}_{; \rho} \Psi_{(1)}^{\rho},
 \end{equation}
and,
\begin{equation}
\frac{D^2 \Psi_{(2)}^{\mu}}{DS^2}= R^{\mu}_{\nu \rho \sigma}\Pi^{\nu} U^{\rho} \Psi_{(2)}^{\sigma}+ \hat{f}^{\mu}_{; \rho} \Psi_{(2)}^{\rho}.
 \end{equation}
 {\bf{Equations (16), (17) are essentially vital  to examine  the stability for different celestial objects in various gravitational fields due to presence of dark matter particles.}}

\subsection{ Equations of Motion of Dipolar Fluid in The Halos}

       {\bf{ The involvement of cosmological constant, a candidate for DE,  has  vital role  to identify the mystery of dark matter. This led Blanchet et al to revisit the description of of dipolar dark matter from particle contents into  fluid-like description [2] . This can be found by replacing $V$ in equation by $W$ the effect of polarization potential, to express interaction of DE on the system .}}

         From this perspective, Blanchet and Le Tiec [4] have postulated that the dynamics of the dipolar fluid in a prescribed gravitational field $g_{\mu \nu}$ is derived from an action of the type found
    \begin{equation}
     S= \int d^{4}x \sqrt{- g}L [ J^{\mu}, \xi^{\mu} \dot{\xi}, g_{\mu \nu}  ]
     \end{equation}
    Provided that the density current $J^{\mu}$  and the polarization vector ${\Pi^{\mu}}$ are new quantities added in dipolar fluids:
     such that:
     {\bf{$ J^{\mu} = \rho U^{\mu}$, and $ \Pi^{\mu} = \rho \xi^{\mu} $ , where $\rho =2mn$ , the inertial mass density to the diploe particles, $n$ the density number of the dipole moment.}} Applying the least action principle on (19)  to  obtain their corresponding set of path equations
     \begin{equation}  \frac{D K^{\mu}}{Ds} = \frac{{f}^{\mu}}{m}  \end{equation}
     and
     $$ \frac{D \Omega}{Ds} = \frac{1}{\hat{\sigma}} \nabla^{\mu} (W-\hat{\Pi} \hat{W}) - R^{\mu}_{\rho \nu \lambda}u^{\rho}\xi^{\nu}K^{\lambda} $$
     where, $\hat{\sigma} = \sqrt(- J^{\mu}J_{\mu})$, $W$ is the density dependent potential, and $K^{\mu}$ is another linear momentum parameterized the dipolar contribution [2] such that
     $$
     K^{\mu} = \frac{P^{\mu}}{2m}.
     $$
     and
     $$
     \hat{\Pi} = \sigma \hat{\pi}
     $$
     where $K^{\mu}$ is the proper time derivative of the linear momentum and$ \hat{\Pi}$ is the density number of the dipole moment.

     The above set of equations can be obtained using its associated Bazanski-Like Lagrangian,
      \begin{equation}
      L = g_{\mu \nu} K^{\mu} \frac{D \Psi_{(1)}^{\nu}}{Ds} +  \Omega_{\mu}\frac{D \Psi_{(2)}^{\nu}}{Ds}+ \bar{f}_{(1)\nu}\Psi_{(1)}^{\nu} +\bar{f}_{(2 )\mu}\Psi_{2}^{\mu},
       \end{equation}
 By taking the variation with respect to  their path deviation vector$\Psi^{\mu}_{(1)}$ and evolution deviation vector $\Psi^{\mu}_{(2)}$ simultaneously. Provided that
 $$ f^{\mu}_{(1)} = \hat{\Pi^{\mu}} \frac{d W}{d \hat{\Pi}}$$
  and $$ \bar{f}^{\mu}_{(2)}=  \frac{1}{\hat{\sigma}} \nabla^{\mu} (W-\hat{\Pi} \hat{W}) - R^{\mu}_{\rho \nu \lambda}u^{\rho}\xi^{\nu}K^{\lambda}. $$

   Thus, using the commutation rule (A.4) and the condition (A.5) we obtain their corresponding path deviation and evolution deviation equations respectively,
      \begin{equation}
     \frac{D^2 \Psi^{\mu}_{(1)}}{Ds^2} = R^{\mu}_{\nu \rho \sigma} K^{\nu} U^{\rho} \Psi^{\sigma}_{1},
     \end{equation}
   and
   \begin{equation}
     \frac{D^2 \Psi^{\mu}_{(2)}}{Ds^2} = R^{\mu}_{\nu \rho \sigma} \Omega^{\nu} U^{\rho} \Psi^{\sigma}_{2} + \bar{f}^{\mu}_{; \rho}\Psi_{2}^{\rho}.
     \end{equation}
{\bf{From equation (22) and (23) , we may also examine the corresponding deviation vectors that are examining the stability of dipolar fluid in the halo due to the presence of DM with taking into consideration the influence of DE.}}

\subsection{Equations of Motion of Fluids in The Accretion Disk}
 Due to the role of non-geodesic equations to explain the behavior of dark matter particles in the accretion disk, as  a collision-less fluid.  We are going to focus on its contribution to mass of the accretion disc and consequently, the accretion process is less efficient than that expected from dissipative fluid ; dark matter gives a significant contribution to the mass of the accretion disk producing an important inflow as in our Galaxy, e.g. a mass growth scaling as $M_{bh} = const. t^{9/16}$ [16].

Thus, we can find out that the equivalence between non-geodesic motions and hydrodynamics flows appears in following two sets of equations
\begin{equation}
\frac{dU^{\alpha}}{ds}+ \Gamma^{\alpha}_{\beta \delta} U^{\beta}U^{\delta} = f^{\alpha},
 \end{equation}
where$ f^{\alpha}$ is described as non-gravitational force, in which its vanishing turns the equation into a geodesic, which becomes

\begin{equation}
\frac{dU^{\alpha}}{ds}+ \Gamma^{\alpha}_{\beta \delta} U^{\beta}U^{\delta} = \frac{1}{E+\hat{P}}h^{\alpha \beta} \hat{P}_{, \beta},
\end{equation}

where , $\hat{P}$ is the pressure of the fluid, $E$ is the over all mass-energy density [7] and $\rho$ is the amount of density .\\
If equation (24) satisfies the first law of thermodynamics.
\begin{equation}
\hat{P}_{, \beta}= \rho c^2 ( \frac{(E + \hat{P})}{\rho c^2})_{, \beta},
 \end{equation}
 then its associated equation of motion of fluids becomes,

\begin{equation}
\frac{dU^{\alpha}}{ds}+ \Gamma^{\alpha}_{\beta \delta} U^{\beta}U^{\delta}= \frac {(\frac{E + \hat{P}}{\rho c^2})_{, \beta}}{ (\frac{E + \hat{P}}{\rho c^2})} h^{\alpha \beta} . \end{equation}

Meanwhile, in case of isobaric pressure, the equation of stream becomes conditionally equivalent to geodesic.
Thus, the appearance of the extra term on the right hand side of equation (4) inspire many authors to interrelate it with  the problem of dark matter as an excess of mass due to the Lagrangian suggested by Kahil and Harko (2009) [1]:

From the above equations, we can find that the excess of mass for a test particle is equivalent to the hydrodynamic equation of motion for a perfect fluid satisfying the first law of thermodynamics. Such an analogy is required to describe the behavior of cluster of fluid circumventing the active galactic nuclei (AGN) it has detected that annihilation of dark matter particles in terms of increase $\gamma$ ray  density in the accretion disc)[6]\\
Accordingly, we  can obtain the hydrodynamic flow of accretion disc  by applying the Euler-Lagrange equation on (2) with taking into account that
\begin{equation}
m(s) {\stackrel{def.}{=}} \frac{(\hat{P}+E)}{\rho c^2}
 \end{equation}
 Using (27), we find  that
\begin{equation} \frac{1}{( E+\hat{P}/ \rho )}\frac{( dE+\hat{P}/ \rho ) }{d \sigma} \approx 2 a_{0}/c^{2}. \end{equation}

 Such a result is inevitable to ensure that the stream of hydrodynamics equations may be expressed with respect to the MOND constant, for  arbitrary parameters $\sigma$ defining the motion.
\section{Dark Matter : Equations of Motion in Bimetric Theories}
 Implementing the concept of geometerization of physics, it is essential to express the motion of non-geodesic equations and their corresponding deviation equation to regulate the behavior of as expressed in particle content or fluid-like in the presence of different bi-metric gravitational fields, able to explain DM at different regions inside spiral galaxies.
\subsection{ Non-Geodesic Trajectories for Bi-gravity }
 Hossenfelder [17] has introduced an alternative version of bi-metric theory, having two different metrics $\bf{g}$ and $\bf{h}$ of Lorentzian signature on a manifold $\bf{M}$  defining the tangential space TM and co-tangential space T*M respectively. These can be obtained in terms of two types of  matter and twin matter; existing individually. Each of them has its own field equations as defined within Riemannian geometry.

  It is well known that implementing bi-gravity theory, without cosmological constants, will be vital to describe motion of dipolar objects in the halos [23]; while the conformal type may be able to describe dark matter as mass excess quantities  found in as in accretion disk circumventing the center  of the Galaxy, as described by strong gravitational fields.\\
 Meanwhile,  theories of bi-metric theories, have one metric combining the two  metrics, with cosmological constant, describing variable speed of light to replace the effect dark energy in big bang scenario [18].\\
 From the previous versions of bi-metric theories [19], we are going to present a generalized form which can be present different types of path and path deviation which can be explained for any bi-metric theory which has two different metrics and curvatures as defined by Riemannian geometry [20]. their Corresponding Lagrangian can be expressed in the following way [21],

\begin{equation}
L{\stackrel{def.}{=}}m_{g}g_{\mu \nu} \Psi_{; \nu} U^{\mu} U^{\nu} + m_{f}f_{\mu \nu} \Phi_{| \nu} V^{\mu} V^{\nu} +(\frac{m_{g}{(s)}_{,\beta}}{m_{g}{(s)}} (g^{\alpha \beta}- U^{\alpha}U^{\beta}))_{;\rho} \Psi^{\rho} + (\frac{m_{f}{(\tau)}_{,\beta}}{m_{f}{(\tau)}} (g^{\alpha \beta}- V^{\alpha}V^{\beta}))_{;\rho} \Psi^{\rho} .
\end{equation}

Thus, regarding\\

{(1)} $ \frac{d \tau}{ds} =0$ , \\
this will give to two separate sets of path equations owing to each parameter by applying the following Bazanski-like Lagrangian:
\begin{equation}
\frac{DU^{\alpha}}{DS}= \frac{m_{(g)}{(s)}_{,\beta}}{m_{(g)}{(s)}} (g^{\alpha \beta}- U^{\alpha}U^{\beta}) ,
\end{equation}
and
\begin{equation}
\frac{DV^{\alpha}}{D \tau}=\frac{m_{(f)}{(\tau)}_{,\beta}}{m_{(f){(\tau})}} (f^{\alpha \beta}- V^{\alpha}V^{\beta}).
\end{equation}
While their corresponding path deviation equations:
\begin{equation}
\frac{D^2\Psi^{\alpha}}{DS^2}= R^{\alpha}_{\beta \gamma \delta} U^{\gamma} U^{\beta} \Psi^{\delta} + (\frac{m_{(g)}{(s)}_{,\beta}}{m_{(g)}} (g^{\alpha \beta}- U^{\alpha}U^{\beta}))_{\rho}\Psi^{\rho},
\end{equation}
And,
\begin{equation}
\frac{D^2\Phi^{\alpha}}{D\tau^2}= S^{\alpha}_{\beta \gamma \delta} V^{\gamma} V^{\beta} \Phi^{\delta} + (\frac{m_{(f)}{(\tau)}_{,\beta}}{m_{(f)}} . (f^{\alpha \beta}- V^{\alpha}V^{\beta}))_{; \rho}\Phi^{\rho},
\end{equation}

{(2)}$ \frac{d \tau}{dS} \neq 0 $   [19], \\

 the two metrics can be related to each other by means of a  quasi-metric one [22].
\begin{equation}
\tilde{g}_{\mu \nu} = g_{\mu \nu} - f_{\mu \nu} + \alpha_{g} ( g_{\mu \nu} - U_{\mu}U_{\nu} ) + \alpha_{f} ( f_{\mu \nu} - V_{\mu}V_{\nu}),
\end{equation}
where $\alpha_{g}$ and $\alpha_{f}$ are arbitrary constants. \\
Such an assumption may give rise to define its related Lagrangian of Bazanski's flavor to describe the geodesic and geodesic deviation equation due to this version of bi-gravity theory.
\begin{equation}
L {\stackrel{def.}{=}} \tilde{g}_{\alpha \beta} U^{\alpha}\frac{\tilde{D} \Psi^{\beta}}{\tilde{D}S},
\end{equation}

$$
  \tilde{\Gamma}^{\alpha}_{\beta \sigma} =  \frac{1}{2}\tilde{g}^{\alpha \delta}( \tilde{g}_{\sigma \delta ,\beta }  +\tilde{g}_{\delta \beta , \sigma } -\tilde{g}_{\beta \sigma ,\delta} ),
$$
 and its corresponding Lagrangian:

 \begin{equation}
 L= \tilde{m(s)} \tilde{g}_{\mu \nu} \tilde{U}^{\mu} ( \frac{d \tilde{\Psi}^{\nu} }{d\tilde{S}} + \tilde{\Gamma}^{\nu}_{\rho \delta} \tilde{\Psi}^{\rho} \tilde{U}^{\delta} )+ \tilde{f_{\mu}}\Psi^{\mu}.
 \end{equation}
 Thus, equation of its path equation can be obtained by taking the variation respect to $\Psi^{\mu}$ to obtain:
 \begin{equation}
\frac{d\tilde{U}^{\alpha}}{d\tilde{S}}+ \tilde{\Gamma}^{\alpha}_{\beta \delta} \tilde{U}^{\beta}\tilde{U}^{\delta} = \frac{{m{\tilde(S)}}_{,\beta}}{{m{\tilde(S)}}} (\tilde{g}^{\alpha \beta}- \tilde{U}^{\alpha}\tilde{U}^{\beta})
, \end{equation}
and using the commutation relation (A.4) and the condition (A.5), we obtain its corresponding deviation equation;
 \begin{equation}
\frac{D^{2}\Psi^{\mu}}{\tilde{DS}^2}= \tilde{R}^{\mu}_{\nu \rho \sigma}\tilde{U}^{\nu}\tilde{U}^{\rho} \tilde{\Psi}^{\sigma} + (\frac{\tilde{m{(\tilde{S})}}_{,\beta}}{m{(\tilde{S})}} (\tilde{g}^{\alpha \beta}- \tilde{U}^{\alpha}\tilde{U}^{\beta}))_{;\rho} \tilde{\Psi}^{\rho}
 ,\end{equation}
 where
 $$
 \tilde{R}^{\alpha}_{.\mu \nu\rho}= \tilde{\Gamma}^{\alpha}_{\mu \rho ,\nu} - \tilde{\Gamma}^{\alpha}_{\mu \nu ,\rho}
 + \tilde{\Gamma}^{\sigma}_{\mu \rho } \tilde{\Gamma}^{\alpha}_{\sigma \rho }  - \tilde{\Gamma}^{\sigma}_{\mu \rho } \tilde{\Gamma}^{\alpha}_{\sigma \rho }.
 $$

\subsection{Equations of Dipolar Moment in Bi-gravity Theory }
Equation of motion of dipolar moment in the presence of bi-metric theory as a candidate to represent DM  as  an interaction between ordinary and twin matter as described by bi-gravity ghost-free theory.

Accordingly, we suggest the following Lagrangian;
\begin{equation}
L {\stackrel{def.}{=}} g_{\alpha \beta} P^{\alpha}\frac{D \Psi_{(1)}^{\beta}}{Ds} + \Omega_{\alpha} \frac{D \Psi_{(2)}^{\beta}}{Ds}   + f_{\alpha}\Psi_{(1)}^{\alpha}+ \hat{f}_{\alpha}\Psi_{(2)}^{\alpha} + f_{\alpha \beta} Q^{\alpha}\frac{D \Phi_{(1)}^{\beta}}{D\tau} + \Delta_{\alpha} \frac{D \Phi_{(2)}^{\beta}}{D\tau}   + k_{\alpha}\Phi_{(1)}^{\alpha}+ \hat{k}_{\alpha}\Psi_{(2)}^{\alpha},
\end{equation}
where, $ Q $  twin matter momentum  vector $ \Delta $ twin matter dipole moment vector ,$ J $ twin non-gravitational force to momentum $  $twin non-gravitational force of dipole moment.
Consequently, taking the variation with respect to $\Psi_{1}$, $\Psi_{2}$ , $\Phi_{1}$ and $\Phi_{2}$  we obtain: the dipolar momentum of ordinary matter, the evolution equation of ordinary matter, the equation of twin dipolar momentum and the equation of twin evolution dipolar moment
\begin{equation}
\frac{D P^{\mu}}{D s} = f^{\mu},
\end{equation}
and its corresponding evolution equation for dipolar moment
\begin{equation}
\frac{D \Omega^{\mu}}{D s} = \hat{f}^{\mu}.
\end{equation}
While, for the twin matter we obtain the equation of its dipolar moment
\begin{equation}
\frac{D Q^{\mu}}{D \tau} = k^{\mu}
\end{equation}
where $k^{\mu}$ is its corresponding non-gravitational force.
Also, the evolution equation of the twin dipolar moment is expressed as follows
\begin{equation}
\frac{D \Delta^{\mu}}{D \tau} = \hat{k}^{\mu},
\end{equation}
in which $k^{\mu}$ is its associate non-gravitational force.

  Moreover, in order to obtain their corresponding deviation equations following the same procedures in for both metrics $g$ and $f$ independently, we get after some manipulations the following set of deviation equations for ordinary matter and twin matter as follows;
for the ordinary matter.
\begin{equation}
\frac{D^2 \Psi_{(1)}^{\mu}}{DS^2}= R^{\mu}_{\nu \rho \sigma}P^{\nu} U^{\rho} \Psi_{(1)}^{\sigma}+ f^{\mu}_{; \rho} \Psi_{(1)}^{\rho},
 \end{equation}
and
\begin{equation}
\frac{D^2 \Psi_{(2)}^{\mu}}{DS^2}= R^{\mu}_{\nu \rho \sigma}\Pi^{\nu} U^{\rho} \Psi_{(2)}^{\sigma}+ \hat{f}^{\mu}_{; \rho} \Psi_{(2)}^{\rho},
 \end{equation}
and for the twin matter
\begin{equation}
\frac{D^2 \Phi_{(1)}^{\mu}}{D\tau^2}= S^{\mu}_{\nu \rho \sigma}Q^{\nu} V^{\rho} \Phi_{(1)}^{\sigma} + k^{\mu}_{; \rho} \Phi_{(1)}^{\rho},
\end{equation}
and
\begin{equation}
\frac{D^2 \Phi_{(2)}^{\mu}}{D\tau^2}= S^{\mu}_{\nu \rho \sigma} {\hat\Pi^{\nu}} V^{\rho} \Phi_{(2)}^{\sigma} + \hat{k}^{\mu}_{; \rho} \Phi_{(2)}^{\rho},
\end{equation}
where, $S^{\alpha}_{\beta \gamma \delta}$ , $V^{\alpha}$, $\hat{\Pi}^{\alpha}$ are their associated curvature, four vector velocity, the polarization vector for particles defined as twin matter respectively.
\subsection{ Dipolar Fluid in Bi-gravity Theory}
Extending the previous ideas as discussed in [3.2], to examine the existence of DM,  using bi-gravity -ghost free theory- to describe both ordinary fluid and twin fluid simultaneously , we suggest the following Lagrangian;
\begin{equation}
 L {\stackrel{def.}{=}} g_{\mu \nu} K^{\mu} \frac{D \Psi_{(1)}^{\nu}}{Ds} +  \Omega_{\mu}\frac{D \Psi_{(2)}^{\nu}}{Ds}+ f_{\mu \nu}\hat{K}^{\mu} \frac{D \hat{\Psi}_{(1)}^{\nu}}{D\tau} +  \hat{\Omega}_{\mu}\frac{D \Psi_{(2)}^{\nu}}{D\tau},
\end{equation}
 where $\hat{K}^{\mu} $ is the twin matter linear momentum and $\Phi_{1}^{\mu}$ its associated deviation vector, $\hat{\Omega}^{\mu}$ the evolution vector associated with twin matter and $\Phi_{2}^{\mu}$ its corresponding deviation vector of the evolution vector for the twin matter,
  in which,   $ \hat{f}_{1}\frac{1}{
     hat{\sigma}} \nabla^{\mu} (W-\hat{\Pi} \hat{W}) - R^{\mu}_{\rho \nu \lambda}u^{\rho}\xi^{\nu}K^{\lambda}.$
Thus, taking the variation with respect to $\Psi_{1}$, $\Psi_{2}$ , $\Phi_{1}$ and $\Phi_{2}$  we obtain
for the ordinary fluid
\begin{equation}
 \frac{D K^{\mu}}{Ds} = f_{1}^{\mu},
 \end{equation}
     and
     \begin{equation}
     \frac{D \Omega^{\mu}}{Ds} = f_{2}^{\mu}. \end{equation}
Also, for the twin fluid
\begin{equation}
 \frac{D \hat{K}^{\mu}}{D\tau} = \hat{f}_{1} ,\end{equation}
     and
      \begin{equation}
      \frac{D \hat{\Omega} }{D\tau} = \frac{1}{\sigma} \nabla^{\mu} (\tilde{W}-\tilde{\Pi} \tilde{W}) - S^{\mu}_{\rho \nu \lambda}V^{\rho}\tilde{\xi}^{\nu}\tilde{K}^{\lambda} , \end{equation}

where $ \tilde{W}^{\mu} $, $\tilde \Pi^{\mu}$, and $K^{\mu}$ are the corresponding twin density dependent potential, the polarization vector, and the related linear momentum vector parameterized due to dipolar description as expressed in bi-gravity theory.
\subsection{Non-Geodesic Equations in AGN: Bimetric theory }
The bi-metric version of equation (4) can be obtained by obtaining the Euler-lagrange equation on the following Lagrangian
\begin{equation}
\tilde{L}= \tilde{g}_{\alpha \beta} \tilde{U}^{\alpha} \frac{D \tilde{\Psi}^{\beta}}{D \tilde{s}}.
\end{equation}
To obtain the corresponding path equation
\begin{equation}
\frac{d\tilde{U}^{\alpha}}{d\tilde{s}}+ \tilde{\Gamma}^{\alpha}_{\beta \delta} \tilde{U}^{\beta}\tilde{U}^{\delta} = \frac{\tilde{m}{(s)}_{,\beta}}{\tilde{m}{(\tilde{s})}} (\tilde{g}^{\alpha \beta}- \tilde{U}^{\alpha}\tilde{U}^{\beta}),
 \end{equation}
and using the commutation relation (A.4) and the condition (A.5), we obtain its corresponding deviation equation;
\begin{equation}
\frac{D^{2}\tilde{\Psi}^{\mu}}{D\tilde{s}^2}= \tilde{R}^{\mu}_{\nu \rho \sigma}\tilde{U}^{\nu}\tilde{U}^{\rho} \tilde{\Psi}^{\sigma} + (\frac{\tilde{m}{(s)}_{,\beta}}{\tilde{{m}\tilde{(s)}}} (\tilde{g}^{\alpha \beta}- \tilde{U}^{\alpha}\tilde{U}^{\beta}))_{;\rho} \tilde{\Psi}^{\rho}.
 \end{equation}

\section{Dark Matter: Problem of Stability }
\subsection{Testing Stability of Celestial Objects by The Geodesic Deviation Vector}
{{\bf{The importance of solving geodesic(non-geodesic) deviation equations that are obtained with its path equation for an object, whether is counted to be a test particle or not is inevitably  used for examining the stability of the system.  The term stability is an analogous meaning to examine the amount of perturbation using deviation vector along its course of motion, to reveal the status of objects in the presence of DM.\\
  In this present work, we are going to implement such a technique which has been applied previously in examining the stability of some cosmological models using two geometric structures [23].

  Recently, this approach has been modified  by [24] to regard the stability condition as a result of by obtaining the scalar value of the deviation vector, independent of any coordinate system  being in covariant form able to study which works for examining the stability problem for any planetary system , and extended for examining the stability of stellar systems orbiting strong gravitational fields [25].}} \\
  Thus, from geodesic deviation equation (11) has its solution expressed in the following manner:
 $$
 \Psi^{\mu} = f(S) C^{\mu},
 $$

  where $ C^{\alpha}$ are constants and $f(S)$ is a function known from the metric. If $ f(S) \rightarrow \infty$ , the system becomes unstable otherwise it is stable. ,
   in a given interval [a,b] in which $\Psi^{\alpha}(S)$ behave monotonically. These quantities can become sensors for measuring the stability of the system are
 \begin{equation}  q~~ {\stackrel{def.}{=}}~~ \lim_{s \rightarrow b} \sqrt{\Psi^{\alpha}\Psi_{\alpha}} . \end{equation}   If  $$q \rightarrow \infty$$  then the system is unstable, otherwise it is always stable. \\
Yet this condition cannot be solely satisfied if one study the case of dipolar particles(fields) .\\
 {\underline{The necessary  and sufficient conditions}} should be related to  the solution of geodesic (non-geodesic) and evolution deviation equations simultaneously  i.e.
$$
 \Psi_{1}^{\mu} = f(S) C_{1}^{\mu},
$$
 and
$$
 \Psi_{2}^{\mu} = f(S) C_{2}^{\mu},
$$

  where $ C_{1}^{\alpha}$, $ C_{2}^{\alpha}$ are constants and $f(S)$ is a function known from the metric. If $ f(S) \rightarrow \infty$ , the system becomes unstable otherwise it is stable. ,
   in a given interval [a,b] in which $\Psi_{1}^{\alpha}(S)$ and $\Psi_{2}^{\alpha}(S)$  behave monotonically. These quantities can become sensors for measuring the stability of the system.\\
   Yet, these conditions can be extended in case of bi-metric theory to be regarded in the following way:

  In case $$ \frac{d \tau}{d s} \neq 0$$  The solution of the set deviation equations (21) and (22)  are

\begin{equation}
\Psi_{(1)}{a}^{\alpha}  = \hat{C}_{1}^{\alpha}f(s),
\end{equation}
and,
\begin{equation}
\Phi_{(1)}{a}^{\alpha}  = \hat{C}_{1}^{\alpha}f(\tau).
\end{equation}
 Thus, we must  obtain two  stability conditions in the following way:
    \begin{equation}  q_{1}~~ {\stackrel{def.}{=}}~~ \lim_{s \rightarrow b} \sqrt{\Psi_{1}^{\alpha}\Psi_{(1) \alpha}} . \end{equation}
     and
     \begin{equation}  {q}_{2}~~ {\stackrel{def.}{=}}~~ \lim_{\tau \rightarrow b} \sqrt{\Phi_{1}^{\alpha}\Phi_{(1)\alpha}} . \end{equation}
Meanwhile, in case of dioplar particles in bimetric metric we get another two more conditions to become:
 \begin{equation}
\Psi_{(2)}{a}^{\alpha}  = \hat{C}_{2}^{\alpha}f(s),
\end{equation}
and,
\begin{equation}
\Phi_{(2)}{a}^{\alpha}  = \hat{C}_{2}^{\alpha}f(\tau).
\end{equation}
Accordingly, in case of the Verozub bi-metric version [9], $ \frac{d \tau}{ds} =0 $, the above conditions appeared for stability  for a test particle and a dipole particle will be reduced to from two to one and from four to two respectively.

\section*{Discussion and Conclusion}
Dark Matter maybe regarded either as a particle or a fluid due to its detection from the source of the gravitational  field.  This has led many authors to revisit its notation and to offer alternatives such as dipolar particles or fluids, an effect of the scalar field and its additional  gravitational  field or even as a result of the projection of higher dimensions  upon other components.
Due to the variety of its differing definitions or notation,  a class of bimetric theories of gravity have been presented to describe the status of these gravitational  fields, whether it is very strong as in the core of the galaxy or a neutron star or weak ones like the Sun that still satisfy the tests of relativity.
This type of theory consists of studying the motion of particles  in terms of their path and deviations vectors. The use of deviation equations is to demonstrate a schematic  approach for estimating the stability of these systems  in a covariant form as mentioned in section 4.  It has been demonstrated that two conditions are essential to examine the stability of a test particle in the presence  of the bimetric  theory.  As these two conditions apply a doubled effect is examined  in the case of their counterparts in bigravity theories.  However applying the Verozub version of bimetric gravity shows its behavior to be the same as the GR.  Owing to  the equation of motion, it is vital to examine the stability of these regions, by solving the geodesic deviation equations, due to inter-relation between geodesic deviation equation and stability conditions.

In our present work, it has been found that non-geodesic equations, as described in bi-metric theory of gravity, may be regarded as a good representative to DM at different regions [26-29].

  Nevertheless, DM has another rival explanation to be examined nearby active galactic nuclei such as SgrA*, due to the excess of mass appeared in equations of relativistic hydrodynamics (27), which is present as a non-geodesic equations equation (3). Also, we have connected between MOND parameters and the rate of mass excess term ,upon parametrization, as shown in equations (6) and (30).

{\bf{Finally, we sum up that the quest of identifying precisely the nature of DM is still under debate. Yet, some authors believe that it may be regarded as a massive neutrino,a super-symmetric neutralino  or even an  axion [30].
The problem of motion as described in  the Riemanian geometry will be extended to be explained by different geometries, admitting non vanishing curvature and torsion simultaneously.\\
Our future work will continue to emphasize the concept of the geometrization of physics in determining the existence  of DM and DE by different classes of Non-Riemaiann geometry, as a further step in demystifying the various notations of both DM and DE.
}}

\section*{ Acknowledgment}  {The author would like to thank Mr. Andrew Gordon for his comments.}

\section*{References}

 {1.} M. E. Kahil, and T. Harko, Mod. Phys. Lett. {\bf{A24}},667(2009). \\
 {2.}P. Wesson, J. Math Phys, {\bf{43}},2423 (2002).\\
  {3.}L. Blanchet , Class. Quant Grav.,{\bf{24}},3541 (2007)\\
 {4.} L. Blanchet and A. Le Tiec, Phys. Rev. D{\bf{78}},024031 (2008).\\
 {\bf{{5.} A. G. Riess et al, Astron. J. {\bf{116}},1009 (1998)}}.\\
 {6.}  B.C. Bromly, ApJs {\bf{197}},2 (2011). \\
 {7.} K. Kleids and N. K. Spyrou, Class. Quant. Grav. {\bf{17}},2965 (2000). \\
{8.}  L. Iorio, Galaxies {\bf {1}},6 (2013). \\
{9.} L. V. Verozub, {\it{Space-time  Relativity and Gravitation, Lambert, Academic Publishing}}(2015).\\
  {10.} S. L. Bazanski,  J. Math. Phys., {\bf {30}},1018 (1989). \\
 {11.} M. Milgrom, Astrophys. J. {\bf{270}},365 (1983). \\
 {12.} M. Milgrom,  Phys. Rev. D{\bf{89}},024027 (2014).\\
 {13.} L. Blanchet, Class. Quant. Grav. {\bf{24}},3541 (2007).\\
 {14.} M. E. Kahil, J. Math. Physics {\bf {47}},052501 (2006). \\
{15.} M. Heydrai-Fard, M.  Mohseni, and H. R. Sepanigi, Phys. Lett. B{\bf{626}}, 230 (2005). \\
{16.} S. Peirani and  J. A. de Freitas Pacheco, Phys. Rev. D{\bf{78}},024031 (2008)\\
{17.} S. Hossenfelder, Phys. Rev. D. {\bf{78}}, 044015 (2008).\\
{18.} J. W. Moffat, Int. J. Mod. Phys. {\bf{A 20}}, 1105 (2005). \\
{19.} Y. Akrami, T. Kovisto, and A. R. Solomon, Gen. Realt. Gravit. {\bf{47}},1838 {(2015)}.\\
 {20.} K. Aoki  and  K. Maeda, Phys. Rev. D{\bf{90}}, 124089 (2014). \\
 {21.} Magd E. Kahil, Gravit. Cosmol.,{\bf{23}}, 70 (2017). \\
 {22.} J. D. Bekenstein, Phys. Rev. D {\bf{48}},3641 (1993). \\
{23.} M.I. Wanas, and M.A Bakry, Proc. MG XI, Part C, 2131(2008). \\ 
 {24.} M.I. Wanas and M.A. Bakry, Astrophys. Space Sci., {\bf{228}},239 (1995). \\
 {25.} Magd E. Kahil,   Odessa Astronomical Publications, {\bf{28/2}}, 126 (2015) \\
 {26.} N. Rosen, Gen. Relativ.  and Gravit., {\bf{4}},  435 (1973). \\
 {27.} S.F. Hassan  and Rachel A. Rosen , JHEP, 126  (2012) \\
{28.} L .Blanchet and L. Heisenberg, Phys Rev. D {\bf{96}},083512 (2017)\\
{29.} L .Blanchet  L. Heisenberg , NORDITA-2015-38 (2015) \\
{30.} I. Pestov,  Proceedings of 5th International Workshop on Complex Structures and Vector Fields, ed S. Dimiev, and K.Sekigawa, World Scientific Singpore, 180 (2001). \\
{31.} Magd E. Kahil Grav. Cosmol., {\bf(24)}, 83 (2018)  \\
{32.} M. Roshan, Phys.Rev. D{\bf{87}}, 044005 (2013). \\

\section*{Appendix (A) }
\subsection*{The Papapertrou Equation in General Relativity: Lagrangian Formalism}
  It is well known that equation of spinning objects in the presence of gravitational field have been studied extensively. This led us to suggest its corresponding  Lagrangian formalism , using a modified Bazanski Lagrangian [31], for  a spinning and precessing object and their corresponding deviation equation in Riemanian geometry in the following way
$$
 L= g_{\alpha \beta} P^{\alpha} \frac{D \Psi^{\beta}}{Ds} + S_{\alpha \beta}\ \frac{D \Psi^{\alpha \beta}}{Ds}+ F_{\alpha}\Psi^{\alpha}+ M_{\alpha \beta}\Psi^{\alpha \beta} \eqno{(A.1)}
 $$
 where
 $ P^{\alpha}= m U^{\alpha}+ U_{\beta} \frac{D S^{\alpha \beta}}{DS}$ and $\Psi^{\mu \nu}$ is the spin deviation tensor.\\
 Taking the variation with respect to $ \Psi^{\mu}$ and $\Psi^{\mu \nu}$ simultaneously we obtain
 $$
\frac{DP^{\mu}}{DS}= F^{\mu},\eqno{(A.2)}
 $$
$$
\frac{DS^{\mu \nu}}{DS}= M^{\mu \nu} \eqno{(A.3)} ,
$$
 where $P^{\mu}$ is the momentum vector, $ F^{\mu} = \frac{1}{2} R^{\mu}_{\nu \rho \delta} S^{\rho \delta} U^{\nu},$ and $R^{\alpha}_{\beta \rho \sigma}$ is the Riemann curvature, $\frac{D}{Ds}$ is the covariant derivative with respect  to a parameter $S$,$S^{\alpha \beta}$ is the spin tensor, $ M^{\mu \nu} =P^{\mu}U^{\nu}- P^{\nu}U^{\mu}$, and $U^{\alpha}= \frac{d x^{\alpha}}{ds}$ is the unit tangent vector to the geodesic. \\
 Using the following identity on both equations (1) and (2)
  $$
  A^{\mu}_{; \nu \rho} - A^{\mu}_{; \rho \nu} = R^{\mu}_{\beta \nu \rho} A^{\beta}, \eqno{(A.4)}
  $$
  where $A^{\mu}$ is an arbitrary vector. \\
 Multiplying both sides with arbitrary vectors, $U^{\rho} \Psi^{\nu}$ as well as using the following condition [15]. \\
 $$
 U^{\alpha}_{; \rho} \Psi^{\rho} =  \Psi^{\alpha}_{; \rho } U^{\rho}, \eqno{(A.5)}
 $$
and $\Psi^{\alpha}$ is its deviation vector associated to the  unit vector tangent $U^{\alpha}$.
 Also in a similar way:
$$
 S^{\alpha \beta}_{; \rho} \Psi^{\rho} =  \Psi^{\alpha \beta}_{; \rho } U^{\rho}, \eqno{(A.6)}
$$

 one obtains the corresponding deviation equations [32]
$$
\frac{D^2 \Psi^{\mu}}{DS^2}= R^{\mu}_{\nu \rho \sigma}P^{\nu} U^{\rho} \Psi^{\sigma}+ F^{\mu}_{; \rho} \Psi^{\rho}, \eqno{(A.7)}
 $$
and
$$
\frac{D^2\Psi^{\mu \nu}}{DS^2}=  S^{\rho [ \mu} R^{\nu ]}_{\rho \sigma \epsilon} U^{\sigma} \Psi^{\epsilon} + M^{\mu \nu}_{; \rho} \Psi^{\rho}.\eqno{(A.8)}
$$

\end{document}